\documentclass[aps,prb,twocolumn,superscriptaddress]{revtex4}
\usepackage{amsmath}
\usepackage{amssymb}
\usepackage[all,arc,knot,web]{xy}
\usepackage{graphicx}
\usepackage{mathtools}
\usepackage{hyperref}
\usepackage{color}
\usepackage{xcolor}
\usepackage{tikz}
\usepackage{wrapfig}
\usepackage{bbm}

\usetikzlibrary{arrows,matrix,decorations.markings}
\usetikzlibrary{positioning}
\usetikzlibrary{decorations.markings}
\usetikzlibrary{decorations.pathmorphing}
\tikzset{->-/.style={decoration={markings,mark=at position #1 with {\arrow{>}}},postaction={decorate}}}
\tikzset{-<-/.style={decoration={markings,mark=at position #1 with {\arrow{<}}},postaction={decorate}}}

\newcommand{\unit}{\mathbf{1}}
\newcommand{\Z}{\mathbb{Z}}
\newcommand{\J}{\mathcal{J}}
\newcommand{\ket}[1]{\left|{#1}\right\rangle}

  \newcommand{\TriangleYpart}[6]{
  	\def\jjA{#1}  	\def\jjB{#2}  	\def\jjC{#3}  	\def\jjD{#4}  	\def\jjE{#5}  	\def\jjF{#6}
  	\TriangleYpartExtended
  }

   \newcommand{\TriangleYpartExtended}[6]{
   	\begin{matrix}\xy
   	0;/r0.12pc/:;
   	(-10,10)*{}="p1";   	(10,10)*{}="p2";   	(0,-10)*{}="p3";   	(-4,5)*{}="p4";   	(4,5)*{}="p5";   	(0,-3)*{}="p6";
   	"p4";"p1" **\dir{-} ?(0.6)*\dir{#5}+(-3,-1) *{\scriptstyle \jjE};
   	"p5";"p2" **\dir{-} ?(0.6)*\dir{#6}+(3.5,-1) *{\scriptstyle \jjF};
   	"p3";"p6" **\dir{-} ?(0.6)*\dir{#1}+(3,-1) *{\scriptstyle \jjA};
   	"p4";"p5" **\dir{-} ?(0.5)*\dir{#4}+(0,3.5) *{\scriptstyle \jjD};
   	"p6";"p5" **\dir{-} ?(0.5)*\dir{#3}+(3.5,0) *{\scriptstyle \jjC};
   	"p6";"p4" **\dir{-} ?(0.5)*\dir{#2}+(-3,0) *{\scriptstyle \jjB};
   	\endxy\end{matrix}
   }
   
   \newcommand{\Boundary}{
   	\begin{matrix}\xy
   	0;<2pt,0pt>:
   	<0pt,2pt>::
   	(-8,40)*\txt{Bulk};
   	 (0,18)*{}="00v18"; (0,22)*{}="00v22"; (0,26)*{}="00v26"; (0,30)*{}="00v30"; (0,34)*{}="00v34"; (0,38)*{}="00v38"; (0,42)*{}="00v42"; (0,46)*{}="00v46"; (0,50)*{}="00v50"; (0,54)*{}="00v54"; (0,58)*{}="00v58"; (0,62)*{}="00v62"; (6,64)*{}="06v64"; 
   	 (6,16)*{}="06v16";
   	  (6,20)*{}="06v20"; (6,24)*{}="06v24"; (6,28)*{}="06v28"; (6,32)*{}="06v32"; (6,36)*{}="06v36"; (6,40)*{}="06v40"; (6,44)*{}="06v44"; (6,48)*{}="06v48"; (6,52)*{}="06v52"; (6,56)*{}="06v56"; (6,60)*{}="06v60"; 
   	 (8,24)*{}="08v24"; (8,26)*{}="08v26"; (8,40)*{}="08v40"; (8,42)*{}="08v42"; (8,56)*{}="08v56"; (8,58)*{}="08v58"; 
   	 (14,20)*{}="14v20"; (14,28)*{}="14v28"; (14,36)*{}="14v36"; (14,44)*{}="14v44"; (14,52)*{}="14v52"; (14,60)*{}="14v60"; 
   	 (16,24)*{}="16v24"; (16,40)*{}="16v40"; (16,56)*{}="16v56"; 
   	 (18,16)*{}="18v16"; (18,32)*{}="18v32"; (18,48)*{}="18v48"; (18,64)*{}="18v64"; (26,16)*{}="26v16"; (26,32)*{}="26v32"; (26,48)*{}="26v48"; (28,18)*{}="28v18"; (28,34)*{}="28v34"; (28,50)*{}="28v50";
   	"28v50"*{\scriptstyle a_1}; "28v34"*{\scriptstyle a_2}; 
   	"08v58"*{\scriptstyle j_1}; "08v42"*{\scriptstyle j_2}; "08v26"*{\scriptstyle j_3}; 
   	"14v60"*{\scriptstyle l_1}; "14v52"*{\scriptstyle l_2}; "14v44"*{\scriptstyle l_3}; "14v36"*{\scriptstyle l_4}; "14v28"*{\scriptstyle l_5}; "14v20"*{\scriptstyle l_6}; 
   	{\ar@{-} "00v18";"06v20"}; {\ar@{-} "00v22";"06v24"}; {\ar@{-} "00v26";"06v28"}; {\ar@{-} "00v30";"06v32"}; {\ar@{-} "00v34";"06v36"}; {\ar@{-} "00v38";"06v40"}; {\ar@{-} "00v42";"06v44"}; {\ar@{-} "00v46";"06v48"}; {\ar@{-} "00v50";"06v52"}; {\ar@{-} "00v54";"06v56"}; {\ar@{-} "00v62";"06v64"}; {\ar@{-} "06v60";"00v58"}; 
   	{\ar@{-}|@{>} "08v24";"16v24"}; {\ar@{-}|@{>} "08v40";"16v40"}; {\ar@{-}|@{>} "08v56";"16v56"}; 
   	{\ar@{-}|@{>} "16v24";"18v32"}; {\ar@{-}|@{>} "16v40";"18v48"}; {\ar@{-}|@{>} "16v56";"18v64"}; 
   	{\ar@{-}|@{>} "18v16";"16v24"}; {\ar@{-}|@{>} "18v32";"16v40"}; 
   	{\ar@{-}|@{>} "18v32";"26v32"}; 
   	{\ar@{-}|@{>} "18v48";"16v56"}; {\ar@{-}|@{>} "18v48";"26v48"};
   	\endxy\end{matrix}
   }
   
   \newcommand{\BoundaryAApart}[4]{
   	\begin{matrix}\xy
   	0;<2pt,0pt>:
   	<0pt,2pt>::
   	(8,40)*{}="08v40";   	(8,42)*{}="08v42";   	(14,28)*{}="14v28";   	(14,36)*{}="14v36";   	(14,44)*{}="14v44";   	(14,52)*{}="14v52";   	(16,24)*{}="16v24";   	(16,40)*{}="16v40";   	(16,56)*{}="16v56";   	(18,32)*{}="18v32";   	(18,48)*{}="18v48";   	(26,32)*{}="26v32";   	(26,48)*{}="26v48";   	(28,34)*{}="28v34";   	(28,50)*{}="28v50";
   	"28v50"*{\scriptstyle #1};   	"28v34"*{\scriptstyle #2};   	"08v42"*{\scriptstyle j_2};   	"14v52"*{\scriptstyle l_2};   	"14v44"*{\scriptstyle #3};   	"14v36"*{\scriptstyle #4};   	"14v28"*{\scriptstyle l_5};
   	{\ar@{-}|@{>} "08v40";"16v40"};   	{\ar@{-}|@{>} "16v24";"18v32"};   	{\ar@{-}|@{>} "16v40";"18v48"};   {\ar@{-}|@{>} "18v32";"16v40"};   	{\ar@{-}|@{>} "18v32";"26v32"};   	{\ar@{-}|@{>} "18v48";"16v56"};   	{\ar@{-}|@{>} "18v48";"26v48"};
   	\endxy\end{matrix}
   }

\newcommand{\TailLabel}[3]{
	\begin{tikzpicture}[scale=0.8]
	\path [lightgray, fill] (1.6,8.8) -- (2.2,8.8) -- (2.2,9.6) -- (2.2,10.4) -- (1.6,10.4) -- cycle;
	\draw [->-=0.5] (2.2,8.8) -- (2.2,9.6);
	\draw [->-=0.5] (2.2,9.6) -- (2.2,10.4);
	\draw [->-=0.5] (2.2,9.6) -- (2.8,9.6);
	\node [] at (2.6,9.4) {\scriptsize $#1$};
	\node [] at (1.8,9.2) {\scriptsize $#2$};
	\node [] at (1.8,10.) {\scriptsize $#3$};
	\end{tikzpicture}
}

\newcommand{\ExpBoundaryBbpAA}{
	\Biggl|\;\begin{matrix}
		\begin{tikzpicture}[scale=0.7]
		\path [lightgray, fill] (1.6,7.2) -- (2.5,7.2) -- (2.7,8.) -- (2.5,8.8) -- (2.7,9.6) -- (2.5,10.4) -- (1.6,10.4) -- cycle;
		\draw [->-=0.5] (1.9,8.8) -- (2.5,8.8);
		\node [] at (2.2,8.58) {\scriptsize $j_{5}$};
		\draw [->-=0.5] (2.7,8.) -- (3.5,8.);
		\node [] at (3.1,7.78) {\scriptsize $a_{1}$};
		\draw [->-=0.5] (2.7,9.6) -- (2.5,10.4);
		\node [] at (2.88,10.07) {\scriptsize $j_{4}$};
		\draw [->-=0.5] (2.7,9.6) -- (3.5,9.6);
		\node [] at (3.1,9.38) {\scriptsize $a_{2}$};
		\draw [->-=0.5] (2.5,8.8) -- (2.7,9.6);
		\node [] at (2.32,9.27) {\scriptsize $j_{3}$};
		\draw [->-=0.5] (2.7,8.) -- (2.5,8.8);
		\node [] at (2.88,8.47) {\scriptsize $j_{2}$};
		\draw [->-=0.5] (2.5,7.2) -- (2.7,8.);
		\node [] at (2.32,7.67) {\scriptsize $j_{1}$};
		\end{tikzpicture}\:
	\end{matrix}\Biggr\rangle
}

\newcommand{\BoundaryBbpAF}{
	\begin{tikzpicture}[scale=0.7]
	\path [lightgray, fill] (1.6,7.2) -- (2.5,7.2) -- (2.7,8.) -- (2.5,8.8) -- (2.7,9.6) -- (2.7,9.6) -- (2.5,10.4) -- (1.6,10.4) -- cycle;
	\draw [->-=0.5] (1.9,8.8) -- (2.5,8.8);
	\node [] at (2.2,8.58) {\scriptsize $j_{5}$};
	\draw [->-=0.5] (2.5,8.8) -- (2.7,9.6);
	\node [] at (2.25,9.29) {\scriptsize $j'_{3}$};
	\draw [->-=0.5] (2.7,9.6) -- (2.5,10.4);
	\node [] at (2.88,10.07) {\scriptsize $j_{4}$};
	\draw [->-=0.5] (2.7,9.6) -- (3.3,9.6);
	\node [] at (3.,9.38) {\scriptsize $a'_{2}$};
	\draw [->-=0.5] (2.5,7.2) -- (2.7,8.);
	\node [] at (2.32,7.67) {\scriptsize $j_{1}$};
	\draw [->-=0.5] (2.7,8.) -- (2.5,8.8);
	\node [] at (2.95,8.49) {\scriptsize $j'_{2}$};
	\draw [->-=0.5] (2.7,8.) -- (3.3,8.);
	\node [] at (3.,7.78) {\scriptsize $a'_{1}$};
	\end{tikzpicture}
}

\newcommand{\bket}[1]{\Biggl|\;{\begin{matrix} #1 \end{matrix}}\;\Biggr\rangle}
\newcommand{\bbra}[1]{\Biggl\langle\;{\begin{matrix} #1 \end{matrix}}\;\Biggr|}
\makeindex

\begin{document} 
	
	\title{Correspondence between bulk entanglement and boundary excitation spectra \\ in 2d gapped topological phases}
	\author{Zhu-Xi Luo}
	\affiliation{Department of Physics and Astronomy, University of Utah, Salt Lake City, Utah, 84112, U.S.A.}
	\author{Brendan G. Pankovich}
	\affiliation{Department of Physics and Astronomy, University of Utah, Salt Lake City, Utah, 84112, U.S.A.}
	\author{Yuting Hu}
	\affiliation{Department of Physics and Institute for Quantum Science and Engineering, Southern University of Science and Technology, Shenzhen 518055, China}
	\affiliation{CAS Key Laboratory of Microscale Magnetic Resonance and Department of Modern Physics, University of Science and Technology of China, Hefei, Anhui 230026, China}
	\author{Yong-Shi Wu}
	\affiliation{State Key Laboratory of Surface Physics, Fudan University, Shanghai 200433, China}
	\affiliation{Department of Physics and Center for Field Theory and Particle Physics, Fudan University, Shanghai 200433, China}
	\affiliation{Collaborative Innovation Center of Advanced Microstructures, Fudan University, Shanghai 200433, China}
	\affiliation{Department of Physics and Astronomy, University of Utah, Salt Lake City, Utah, 84112, U.S.A.}
	\date{\today}
	
	
\begin{abstract}
We study the correspondence between boundary excitation distribution spectrum of non-chiral topological orders on an open surface $\mathcal{M}$ with gapped boundaries and the entanglement spectrum in the bulk of gapped topological orders on a closed surface. The closed surface is bipartitioned into two subsystems, one of which has the same topology as $\mathcal{M}$. Specifically, we focus on the case of generalized string-net models and discuss the cases where $\mathcal{M}$ is a disk or a cylinder. When $\mathcal{M}$ has the topology of a cylinder, different combinations of boundary conditions of the cylinder will correspond to different entanglement cuts on the torus. When both boundaries are charge (smooth) boundaries, the entanglement spectrum can be identified with the boundary excitation distribution spectrum at infinite temperature and constant fugacities. Examples of toric code, $\mathbb{Z}_N$ theories, and the simple non-abelian case of doubled Fibonacci are demonstrated.
\end{abstract}
	
	\maketitle
	
	\section{Introduction}\label{Intro}
	Entanglement spectrum is the spectrum of entanglement Hamiltonian $H_{E}$, defined from the reduced density matrix of a bipartition of the system $\rho_A = e^{-H_E}$. It was introduced ten years ago by Li and Haldane \cite{Spectrum} as an identification of topological order in fractional quantum Hall states. They showed that the state counting of low-lying entanglement spectrum of the model, e.g. those of the Laughlin and the Moore-Read states, is identical to the counting of conformal field theory modes describing low-energy boundary excitations.
	
	Similar correspondence between bulk entanglement and boundary spectra has been studied analytically in various topological phases. For topological insulators, superconductors and general symmetry protected topological phases, degeneracies of bulk entanglement spectrum correspond to gapless edge modes\cite{Pollman, Fidkowski, Turner}.
	For fractional quantum hall systems, rigorous results on a large class of trial wave functions have been obtained \cite{Chandran, Estienne}. In Ref. \cite{Dubail}, it was shown that the boundary conformal field theory (BCFT) and the bulk CFT used to construct the ground state trial wave function are isomorphic up to a Wick rotation. In general (2+1)d topological quantum systems possessing edge states described by a chiral (1+1)d CFT, a cut-and-glue method was applied in Ref. \cite{QiChiral} to show that the reduced density matrix of a subregion in the bulk topological state is a thermal density matrix of the chiral edge state CFT that appear at the spatial boundary of the bulk subregion. Later a geometric proof was proposed \cite{Swingle}.
	
	Above methods can only be applied to chiral topological phases where there are chiral edge states appearing on the boundaries of the system. In Ref.\cite{Vidal}, the $\mathbb{Z}_2$ spin liquid (toric code) model was discussed using free boundary conditions on a cylinder. An exact correspondence was found between the boundary and entanglement spectra. But it is yet to be clear whether and how the three smooth-rough, rough-rough, smooth-smooth gapped boundary conditions (in the sense of \cite{KK}) can be related to the entanglement spectrum in the bulk. 
	We would like to study these possibilities and explore the non-chiral version of the correspondence in generalized string-net models with boundaries.
	
	String-net models \cite{StringNet} describe a large class non-chiral (2+1)d topological phases, including all those whose low-energy effective theories are discrete gauge or doubled Chern-Simons theories. The model was first constructed for closed surfaces but has been generalized to open surfaces for specific cases \cite{Bravyi, Shor} and then generally formulated using module category \cite{KK, Qalgebra}. Recently the explicit boundary Hamiltonian has been worked out using Frobenius algebras \cite{Bdry1,Bdry2}. We will apply this formalism in the remaining of the paper, because it allows for a more convenient way to solve the spectrum and eigenstates.
	
	Entanglement properties of string-net models were first discussed in \cite{WenTEE} on a sphere. An universal constant term subleading to the area law was found and named as topological entanglement entropy. On nontrivial surfaces like a torus, the entanglement entropy turns out to be more complicated. In the case where the bipartition is done by cutting the torus into two cylinders, Ref. \cite{MES} carried out the calculation for toric code model, and subsequently defined the concept of minimally entangled states. Ref. \cite{Luo} generalized them to minimally entangled sectors, which are classes of minimally entangled states that however be superposed, will always give the same entanglement entropy. 
	In this work we focus on the correspondence between (1) boundary excitation spectrum of string-net model on a open surface (2) the entanglement spectrum obtained from bipartitioning a closed surface into two subsystems $A$, $B$ so that subsystem $A$ has the same topology of the open surface where the boundary excitation spectrum is calculated. {By boundary excitation spectrum, we mean not the usual energy spectrum, but the excitation distribution spectrum. The latter is insensitive to the detailed energy dispersion relations and only contains the information of topological quantum numbers, which makes it more robust for generalizations beyond exactly solvable models. It is defined through the following density matrix in the grand canonical ensemble,
	\begin{equation}
	\rho=\oplus_{n,\{n_\alpha\}}\rho_{n,\{n_\alpha\}},~~\rho_{n,\{n_\alpha\}}=e^{-\beta\sum_{\alpha} n_\alpha(\varepsilon_\alpha-\mu_\alpha)}~\mathbbm{1}.
	\end{equation}
   Here $n$ is the total number of excitations in the system, consisting of $n_\alpha$ of $\alpha$-type excitation, $n_\beta$ of $\beta$ type excitation, etc.. $\mathbbm{1}$ is an identity matrix of dimension deg$_{n,\{n_\alpha\}}$, which counts the degeneracy of states given such a distribution of excitations. $\varepsilon_\alpha$ is the energy needed to excite an excitation of type $\alpha$, and $\mu_\alpha$ is the chemical potential. Using the concept of fugacity $z_\alpha=e^{\beta\mu_\alpha}$, we can rewrite the above expression as
   $
   \rho_{n,\{n_\alpha\}}=\prod_\alpha (z_\alpha^{n_\alpha}~e^{-\beta n_\alpha\varepsilon_\alpha})~\mathbbm{1}.
   $
   In the limits of high temperature $\beta\rightarrow0$ and constant fugacity $z_\alpha$, the density matrix further simplifies to
   \begin{equation}
   \rho_{n,\{n_\alpha\}}=(\prod_{\alpha}z_\alpha^{n_\alpha})~\mathbbm{1}.
   \label{eq:DM}
   \end{equation}

We will show that this density matrix matches with the reduced density matrix obtained from entanglement calculations as long as we identify $z_\alpha=d_\alpha$, the ratio between quantum dimension of excitation $\alpha$ and the total quantum dimension of the system. } These results are universal because the string-net model, as a state sum topological field theory, is a fix-point of the topological phase. For general systems that lie in the same topological phase but deviate from the exactly solvable string-net model, the details of the bulk entanglement and the boundary excitation distribution spectra can vary, but the topological information encoded in the spectra remains invariant. 
	
In section \ref{sec:SN}, we review the construction of string-net model on open systems using Frobenius algebras. Section \ref{sec:disk} demonstrates the correspondence in the case of a boundary excitation distribution spectrum for a disk and the entanglement spectrum for a disk-shaped subsystem. Simple examples of the toric code and the doubled Fibonacci models are presented. Then in \ref{sec:cyl} we study the correspondence for boundary excitation distribution spectrum on a cylinder and entanglement spectrum for a torus bipartitioned into a cylindrical subsystem $A$ and the rest. The toric code \cite{Toric} case is discussed in \ref{subsec:EntCyl}. Each of the three different possible boundary conditions of the toric code model on a cylinder correspond to an entanglement spectrum on a torus with different entanglement cuts, made possible through the introduction of minimally entangled sectors. $\mathbb{Z}_N$ models are also demonstrated. Then we generalize to the non-abelian cases in section \ref{subsec:Fib}. Finally in \ref{sec:discussion} we comment on subtleties arising from the most general cases.
	
\section{String-net Model with Boundaries}\label{sec:SN}
	We briefly review the general theory of string-net model on surfaces with boundaries. Typical examples will be presented in the latter sections.
	
	The input data $\left\{I, d, \delta, G\right\}$ in the bulk of string-net models form a unitary fusion category $\mathcal{C}$. 
	The model is defined on a trivalent graph on a closed oriented surface. Degrees of freedom live on links of the graph. 
	For each link, we assign a string type $j\in I=\{j=0,1,...,N\}$, where $I$ is called the label set. In the case of lattice gauge theories, $j$'s label the irreducible representations of a group. More generally, they can label irreducible representations of quantum groups. 
	The Hilbert space is spanned by all configurations of the labels on links. Each label $j$ has a ``conjugate'' $j^*\in I$, satisfying $j^{**}=j$. There is unique  ``vacuum'' label $j=0$ with $0^*=0$. We require the state to be the same if one reverses the direction of one link and replaces the label $j$ by $j^*$, which is a graphical realization of time reversal symmetry. 
	
	We associate to each string type a number $d_j$ called quantum dimension of $j$, and define the total quantum dimension to be $D=\sum_{j\in I} d_j^2$. We further assign to each three string types a tensor $\delta_{ijk}$ which specifies the branching rules of a trivalent graph. If for some $i,j,k\in I$ one has $\delta_{ijk}=1$, then the three string types are allowed to meet at a vertex. Otherwise their meeting is not energetically favored, i.e., we will have charge excitations on the corresponding vertex. (We will focus on the multiplicity-free cases for convenience.)
	
	Given the quantum dimensions and fusion rules, we define the symmetrized $6j$-symbols, often denoted as $G$. They are complex numbers satisfying the following conditions\cite{Full}:
	\begin{align}
	\label{eq:6jcond}
	\begin{array}{ll}
	&G^{ijm}_{kln}=G^{mij}_{nk^{*}l^{*}}
	=G^{klm^{*}}_{ijn^{*}}=\iota_m\iota_n\,\overline{G^{j^*i^*m^*}_{l^*k^*n}},\\
	&\sum_{n}{d_{n}}G^{mlq}_{kp^{*}n}G^{jip}_{mns^{*}}G^{js^{*}n}_{lkr^{*}}
	=G^{jip}_{q^{*}kr^{*}}G^{riq^{*}}_{mls^{*}},\\
	&\sum_{n}{d_{n}}G^{mlq}_{kp^{*}n}G^{l^{*}m^{*}i^{*}}_{pk^{*}n}
	=\frac{\delta_{iq}}{d_{i}}\delta_{mlq}\delta_{k^{*}ip},\\
	\end{array}
	\end{align}
	where the first condition specifies tetrahedral symmetry, the second the pentagon identity, and the third orthogonality condition. The number $\iota_j$ is the Frobenius-Schur indicator. In the example of lattice gauge theories, this indicator tells whether the representation $j$ is real, complex, or pseudoreal. Then $d_j=\iota_j\mathrm{dim}(j)$ with $\mathrm{dim}(j)$ being the corresponding dimension of the space of the representation $j$; and the tensor $G_{kln}^{ijm}$ is the (symmetrized) Racah $6j$ symbol for the group. In this example, string-net model can be mapped to the Kitaev's quantum double model. 
	
	Two types of local operators are needed to specify the Hamiltonian. On every vertex $v$, we have ${Q}_v=\delta_{ijk}$ that acts on the labels of three edges incoming to the vertex $v$. On every plaquette $p$, we have $B_p^s$ with $s\in I$, which acts on the boundary edges of the plaquette $p$. Its matrix elements on a triangular plaquette is \cite{Full},
	\begin{equation}
	\label{eq:Bps::InLW}
	\begin{split}
	\Biggl\langle
	\TriangleYpart{j_4}{j^{\prime}_1}{j^{\prime}_2}{j^{\prime}_3}{j_5}{j_6}{>}{<}{>}{<}{<}{<}
	\Biggr|
	B_p^s
	\Biggl|\TriangleYpart{j_4}{j_1}{j_2}{j_3}{j_5}{j_6}{>}{<}{>}{<}{<}{<}\Biggr\rangle\nonumber
	=
	& v_{j_1}v_{j_2}v_{j_3}v_{j'_1}v_{j'_2}v_{j'_3}\\
	& \times 
	G^{j_5j^*_1j_3}_{sj'_3j^{\prime*}_1}G^{j_4j^*_2j_1}_{sj'_1j^{\prime*}_2}G^{j_6j^*_3j_2}_{sj'_2j^{\prime*}_3},
	\end{split}
	\end{equation}
	where $v_j=\sqrt{d_j}=\frac{1}{G^{j^*j0}_{0\,0\,j}}$. The same pattern for $B_p^s$ applies when the plaquette $p$ is a quadrangle, a pentagon, etc.. 
	
	Defining $B_p=\frac{1}{D}\sum_{s}d_sB_p^{s}$, the operators ${Q}_v$'s and $B_p$'s are mutually-commuting. Furthermore, they are also projectors: ${Q}_v^2={Q}_v$ and $B_p^2=B_p$. The Hamiltonian of the model is
	\begin{equation}
	\label{eq:HamiltonianLW}
	H_{\text{bulk}}=\sum_{v}(1-{Q}_v)+\sum_{p}(1-B_p),  
	\end{equation}
	where the sum runs over vertices $v$ and plaquettes $p$ of the whole trivalent graph. Because of the commutative property of $Q_v$ and $B_p$'s, the Hamiltonian is exactly soluble. Ground states satisfies ${Q}_v=B_p=1$ for all $v$, $p$.
	
	The bulk ground states are invariant under any composition of the following elementary (dual) Pachner moves:
	\newcommand{\ExpTMoveTwoTwo}{
		\Biggl|\;\begin{matrix}
			\begin{tikzpicture}[scale=0.7]
			\draw [->-=0.5] (1.6,8.8) -- (2.,9.6);
			\node [] at (2.06,9.07) {\scriptsize $j_{2}$};
			\draw [->-=0.5] (1.6,10.4) -- (2.,9.6);
			\node [] at (1.54,9.87) {\scriptsize $j_{1}$};
			\draw [->-=0.5] (3.2,8.8) -- (2.8,9.6);
			\node [] at (3.26,9.33) {\scriptsize $j_{3}$};
			\draw [->-=0.5] (3.2,10.4) -- (2.8,9.6);
			\node [] at (2.74,10.13) {\scriptsize $j_{4}$};
			\draw [->-=0.5] (2.8,9.6) -- (2.,9.6);
			\node [] at (2.4,9.82) {\scriptsize $j_{5}$};
			\end{tikzpicture}\:
		\end{matrix}\Biggr\rangle
	}
	\newcommand{\ExpTMoveTwoTwoAA}{
		\displaystyle\sum_{j'_{5}}G^{j_{1} j_{2} j_{5}}_{j_{3} j_{4} j'_{5}}\mathrm{v}_{j_{5}}\mathrm{v}_{j'_{5}}
		\Biggl|\;\begin{matrix}
			\begin{tikzpicture}[scale=0.7]
			\draw [->-=0.5] (1.6,8.8) -- (2.4,9.2);
			\node [] at (2.11,8.77) {\scriptsize $j_{2}$};
			\draw [->-=0.5] (1.6,10.4) -- (2.4,10.);
			\node [] at (1.89,9.97) {\scriptsize $j_{1}$};
			\draw [->-=0.5] (2.4,9.2) -- (2.4,10.);
			\node [] at (2.77,9.6) {\scriptsize $j'_{5}$};
			\draw [->-=0.5] (3.2,10.4) -- (2.4,10.);
			\node [] at (2.91,9.97) {\scriptsize $j_{4}$};
			\draw [->-=0.5] (3.2,8.8) -- (2.4,9.2);
			\node [] at (2.69,8.77) {\scriptsize $j_{3}$};
			\end{tikzpicture}\:
		\end{matrix}\Biggr\rangle
	}
	\newcommand{\ExpTMoveOneThree}{
		\Biggl|\;\begin{matrix}
			\begin{tikzpicture}[scale=0.7]
			\draw [->-=0.5] (1.6,10.4) -- (2.2,9.8);
			\node [] at (1.71,9.91) {\scriptsize $j_{1}$};
			\draw [->-=0.5] (2.2,9.) -- (2.2,9.8);
			\node [] at (2.49,9.4) {\scriptsize $j_{2}$};
			\draw [->-=0.5] (2.8,10.4) -- (2.2,9.8);
			\node [] at (2.69,9.91) {\scriptsize $j_{3}$};
			\end{tikzpicture}\:
		\end{matrix}\Biggr\rangle
	}
	\newcommand{\ExpTMoveOneThreeAA}{
		\displaystyle\sum_{j_{4},j_{5},j_{6},}\frac{\mathrm{v}_{j_{5}}\mathrm{v}_{j_{6}}\mathrm{v}_{j_{4}}}{\sqrt{D}}G^{j_{1} j_{2} j_{3}}_{j_{5} j_{6} j_{4}}
		\Biggl|\;\begin{matrix}
			\begin{tikzpicture}[scale=0.7]
			\draw [->-=0.5] (1.6,10.4) -- (1.9,10.1);
			\node [] at (1.56,10.06) {\scriptsize $j_{1}$};
			\draw [->-=0.5] (2.3,9.) -- (2.3,9.4);
			\node [] at (2.59,9.2) {\scriptsize $j_{2}$};
			\draw [->-=0.5] (2.3,9.4) -- (1.9,10.1);
			\node [] at (1.85,9.61) {\scriptsize $j_{4}$};
			\draw [->-=0.5] (2.7,10.1) -- (1.9,10.1);
			\node [] at (2.3,10.32) {\scriptsize $j_{6}$};
			\draw [->-=0.5] (3.,10.4) -- (2.7,10.1);
			\node [] at (3.04,10.06) {\scriptsize $j_{3}$};
			\draw [->-=0.5] (2.7,10.1) -- (2.3,9.4);
			\node [] at (2.75,9.61) {\scriptsize $j_{5}$};
			\end{tikzpicture}\:
		\end{matrix}\Biggr\rangle
	}
	\newcommand{\ExpTMoveThreeOne}{
		\Biggl|\;\begin{matrix}
			\begin{tikzpicture}[scale=0.7]
			\draw [->-=0.5] (1.6,10.4) -- (1.9,10.1);
			\node [] at (1.56,10.06) {\scriptsize $j_{1}$};
			\draw [->-=0.5] (2.3,9.) -- (2.3,9.4);
			\node [] at (2.59,9.2) {\scriptsize $j_{2}$};
			\draw [->-=0.5] (2.3,9.4) -- (1.9,10.1);
			\node [] at (1.85,9.61) {\scriptsize $j_{4}$};
			\draw [->-=0.5] (2.7,10.1) -- (1.9,10.1);
			\node [] at (2.3,10.32) {\scriptsize $j_{6}$};
			\draw [->-=0.5] (3.,10.4) -- (2.7,10.1);
			\node [] at (3.04,10.06) {\scriptsize $j_{3}$};
			\draw [->-=0.5] (2.7,10.1) -- (2.3,9.4);
			\node [] at (2.75,9.61) {\scriptsize $j_{5}$};
			\end{tikzpicture}\:
		\end{matrix}\Biggr\rangle
	}
	\newcommand{\ExpTMoveThreeOneAA}{
		\frac{\mathrm{v}_{j_{5}}\mathrm{v}_{j_{4}}\mathrm{v}    _{j_{6}}}{\sqrt{D}}G^{j_{1}^* j_{3}^* j_{2}^*}_{j_{5} j_{4}^* j_{6}^*}
		\Biggl|\;\begin{matrix}
			\begin{tikzpicture}[scale=0.7]
			\draw [->-=0.5] (1.6,10.4) -- (2.3,9.8);
			\node [] at (1.78,9.9) {\scriptsize $j_{1}$};
			\draw [->-=0.5] (2.3,9.) -- (2.3,9.8);
			\node [] at (2.59,9.4) {\scriptsize $j_{2}$};
			\draw [->-=0.5] (3.,10.4) -- (2.3,9.8);
			\node [] at (2.82,9.9) {\scriptsize $j_{3}$};
			\end{tikzpicture}\:
		\end{matrix}\Biggr\rangle
	}
	\begin{equation}
	\label{eq:PachnerBulk}
	\begin{split}
	&T_{2\rightarrow 2}\ExpTMoveTwoTwo
	=\ExpTMoveTwoTwoAA\nonumber\\
	&T_{1\rightarrow 3}\ExpTMoveOneThree
	=\ExpTMoveOneThreeAA\nonumber\\
	&T_{3\rightarrow 1}\ExpTMoveThreeOne
	=\ExpTMoveThreeOneAA \\
	\end{split}
	\end{equation}
	
	The boundary theory of string-net models was first formulated in an abstract language by Ref.\cite{KK,Qalgebra}, building on the module category $\mathcal{M}$ of the input category $\mathcal{C}$.  An alternative was developed in Ref.\cite{Bdry1, Bdry2}, where the Hamiltonian is written explicitly in terms of input data and can be used to solve for the spectrum and eigenstates. The basic object in this formulation is a separable Frobenius algebra $\mathcal{A}$ constructed from $\mathcal{C}$ and the boundary degrees of freedom form modules of the algebra. These two formulations \cite{KK, Bdry1} are mathematically equivalent due to a theorem: the category of right modules over an algebra $\mathcal{A}$ is equivalent to the right module category $\mathcal{M}$ over (unitary fusion) $\mathcal{C}$ \cite{Catbook}.
	
\begin{figure}[h]
		\centering
		\includegraphics[]{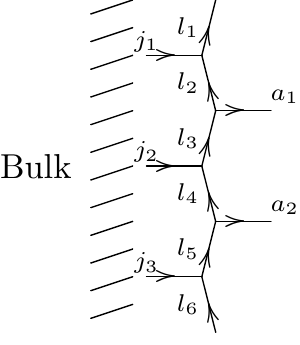}
		\caption{Boundary is a wall carrying open links. Degrees of freedom on the wall are labeled by $l\in I$, and open links by $a\in I_{\mathcal{A}}$.}
		\label{fig:Boundary}
	\end{figure}

	A Frobenius algebra is the subset $I_\mathcal{A}\subset I$ equipped with a multiplication structure $f_{ijk}$ which describes the fusion of open links $i\otimes j\rightarrow k^*$ and satisfies the following constraints:
	\begin{equation}
	\begin{split}
	& \text{Association:}~ \sum_c f_{abc^*} f_{cde^*} G_{de^*g}^{abc^*}v_cv_g=f_{age^*}f_{bdg^*},\\
	& \text{Non-degeneracy}~ f_{aa^*0}\neq 0, \forall a\in I_{\mathcal{A}}.\\
	\end{split}
	\end{equation}
	We choose to normalize as $f_{aa^*0}=1 \forall a\in I_{\mathcal{A}}$. 
	There are two types of boundary degrees of freedom: $l\in I$ on the wall and $a\in I_\mathcal{A}$ on the open links, as indicated in Fig.\ref{fig:Boundary}. 
	
	The boundary Hamiltonian can then be defined using the Frobenius algebra:
	\begin{equation}
	H_{\text{bdry}}=\sum_n (1-\overline{Q}_n)+ \sum_n (1-\overline{B}_n),
	\end{equation}
	with $\overline{Q}_n$ acting on the open link $n$ and projecting the boundary degrees of freedom to $I_\mathcal{A}$:
	\begin{equation}\label{eq:Qn}
	\overline{Q}_n\bket{\TailLabel{a_n}{j_1}{j_2}}=\delta_{a_n\in I_\mathcal{A}}\bket{\TailLabel{a_n}{j_1}{j_2}}.
	\end{equation}
	$\overline{B}_n$ is a combination of $\overline{B}^t_n$'s,
	\begin{equation}
	\overline{B}_n=\frac{1}{d_\mathcal{A}}\sum_{t\in I_\mathcal{A}} \overline{B}^t_n,\ \ d_\mathcal{A}=\sum_{t\in I_\mathcal{A}} d_t.
	\end{equation}
	The operator $\overline{B}^t_p$ fuses a string $t$ to the boundary ``half plaquette'' as follows:
	\begin{equation}
	\begin{split}
	\overline{B}^t_{n}: \ExpBoundaryBbpAA
	\mapsto
	\sum_{a'_{1},a'_{2},j'_{2},j'_{3}}
	f_{t^* {a'_{2}}^* a_{2}} f_{a_{1} t {a'_{1}}^*}u_{a_{1}}
	u_{a_{2}}u_{a'_{1}}\\
	\times u_{a'_{2}}
	G^{j_{4}^* j_{3} a_{2}^*}_{t^* {a'_{2}}^* j'_{3}}
	G^{j_{5} j_{2} j_{3}^*}_{t^* {j'_{3}}^* j'_{2}}
	G^{t^* {j'_{2}}^* j_{2}}_{j_{1} a_{1}^* a'_{1}}
	v_{j_{2}}v_{j_{3}}v_{j'_{2}}v_{j'_{3}}
	\bket{\BoundaryBbpAF}.
	\end{split}
	\end{equation}
	One can easily check that the boundary plaquette operators are mutual commuting projection operators and they commute with the bulk operators. The full Hamiltonian of the system will then be
	\begin{equation}
	\label{eq:FullHam}
	H=H_{\text{bulk}}+\epsilon H_{\text{bdry}},
	\end{equation}
	with $\epsilon$ a positive number.
	
	Similar to the elementary Pachner moves in the bulk \eqref{eq:PachnerBulk}, one can use the Frobenius algebra $A$ to define transformations associated with on the boundaries of a graph. The ground states of string-net models with boundaries are invariant under the following elementary moves:
	\newcommand{\ExpTBdryMoveOneTwo}{
		\Biggl|\;\begin{matrix}
			\begin{tikzpicture}[scale=1]
			\path [lightgray, fill] (1.6,8.8) -- (2.2,8.8) -- (2.2,9.6) -- (2.2,10.4) -- (1.6,10.4) -- cycle;
			\draw [->-=0.5] (2.2,9.6) -- (2.8,9.6);
			\node [] at (2.5,9.38) {\scriptsize $a_{1}$};
			\draw [->-=0.5] (2.2,8.8) -- (2.2,9.6);
			\node [] at (1.98,9.2) {\scriptsize $i$};
			\draw [->-=0.5] (2.2,9.6) -- (2.2,10.4);
			\node [] at (1.98,10.) {\scriptsize $j$};
			\end{tikzpicture}\:
		\end{matrix}\Biggr\rangle
	}
	\newcommand{\ExpTBdryMoveOneTwoAC}{
		\displaystyle\sum_{a_{2},a_{3}}\frac{\mathrm{u}_{a_{1}}\mathrm{u}_{a_{2}}\mathrm{u}_{a_{3}}}{\sqrt{\mathrm{d}_A}}\displaystyle\sum_{k}\mathrm{v}_{k}f_{a_{2}^* a_{3}^* a_{1}}G^{j^* i a_{1}^*}_{a_{2}^* a_{3}^* k}
		\Biggl|\;\begin{matrix}
			\begin{tikzpicture}[scale=1]
			\path [lightgray, fill] (1.6,8.8) -- (2.2,8.8) -- (2.2,9.6) -- (2.2,10.4) -- (1.6,10.4) -- cycle;
			\draw [->-=0.5] (2.2,8.8) -- (2.2,9.3);
			\node [] at (1.98,9.05) {\scriptsize $i$};
			\draw [->-=0.5] (2.2,9.3) -- (2.2,9.9);
			\node [] at (1.98,9.6) {\scriptsize $k$};
			\draw [->-=0.5] (2.2,9.3) -- (3.,9.3);
			\node [] at (2.6,9.08) {\scriptsize $a_{2}$};
			\draw [->-=0.5] (2.2,9.9) -- (2.2,10.4);
			\node [] at (1.98,10.15) {\scriptsize $j$};
			\draw [->-=0.5] (2.2,9.9) -- (3.,9.9);
			\node [] at (2.6,10.12) {\scriptsize $a_{3}$};
			\end{tikzpicture}\:
		\end{matrix}\Biggr\rangle
	}
	\newcommand{\ExpTBdryMoveTwoOne}{
		\Biggl|\;\begin{matrix}
			\begin{tikzpicture}[scale=1]
			\path [lightgray, fill] (1.6,8.8) -- (2.2,8.8) -- (2.2,9.6) -- (2.2,10.4) -- (1.6,10.4) -- cycle;
			\draw [->-=0.5] (2.2,8.8) -- (2.2,9.3);
			\node [] at (1.98,9.05) {\scriptsize $i$};
			\draw [->-=0.5] (2.2,9.3) -- (2.2,9.9);
			\node [] at (1.98,9.6) {\scriptsize $k$};
			\draw [->-=0.5] (2.2,9.3) -- (3.,9.3);
			\node [] at (2.6,9.08) {\scriptsize $a_{2}$};
			\draw [->-=0.5] (2.2,9.9) -- (2.2,10.4);
			\node [] at (1.98,10.15) {\scriptsize $j$};
			\draw [->-=0.5] (2.2,9.9) -- (3.,9.9);
			\node [] at (2.6,10.12) {\scriptsize $a_{3}$};
			\end{tikzpicture}\:
		\end{matrix}\Biggr\rangle
	}
	\newcommand{\ExpTBdryMoveTwoOneAC}{
		\displaystyle\sum_{a_{1}}\frac{\mathrm{u}_{a_{1}}\mathrm{u}_{a_{2}}\mathrm{u}_{a_{3}}}{\sqrt{\mathrm{d}_A}}\mathrm{v}_{k}f_{a_{2} a_{3} a_{1}^*}G^{j a_{1} i^*}_{a_{2}^* k^* a_{3}}
		\Biggl|\;\begin{matrix}
			\begin{tikzpicture}[scale=1]
			\path [lightgray, fill] (1.6,8.8) -- (2.2,8.8) -- (2.2,9.6) -- (2.2,10.4) -- (1.6,10.4) -- cycle;
			\draw [->-=0.5] (2.2,8.8) -- (2.2,9.6);
			\node [] at (1.98,9.2) {\scriptsize $i$};
			\draw [->-=0.5] (2.2,9.6) -- (2.2,10.4);
			\node [] at (1.98,10.) {\scriptsize $j$};
			\draw [->-=0.5] (2.2,9.6) -- (3.,9.6);
			\node [] at (2.6,9.38) {\scriptsize $a_{1}$};
			\end{tikzpicture}\:
		\end{matrix}\Biggr\rangle
	}
	\begin{small}
		\begin{equation}
		\label{eq:PachnerBdry}
		\begin{split}
		&T_{1\rightarrow 2} \ExpTBdryMoveOneTwo\\
		=&\ExpTBdryMoveOneTwoAC,\nonumber\\
		&T_{2\rightarrow 1} \ExpTBdryMoveTwoOne\\
		=&\ExpTBdryMoveTwoOneAC.
		\end{split}
		\end{equation} 
		where $u_a=\sqrt{v_a}$.
	\end{small}
	
For any input data in the bulk, there is always a trivial Frobenius algebra $\mathcal{A}_0$ corresponding to $I_\mathcal{A}=\{0\}$. This is often called the ``smooth'' boundary in literature, but we will use instead the term ``charge boundary'' instead because in this case the boundary Hamiltonian reduces to the charge term only,
$H_{\text{bdry}}^{\text{charge}}=-\sum_n (1-\overline{Q}_n).$
On the other hand, the ``rough'' or ``flux'' boundaries $I_\mathcal{A}=I$ are not guaranteed to exist for general string-net models. However, they do appear in important examples like the toric code and the doubled Fibonacci model, which will be discussed in latter sections. When exist, the Hamiltonian is simply
$H_{\text{bdry}}^{\text{flux}}=-\sum_n (1-\overline{B}_n).$
	
	\newcommand{\BpBasisEff}[5]{
		\begin{tikzpicture}[scale=0.8]
		\path [lightgray, fill] (1.6,8.) -- (2.4,8.) -- (2.4,10.4) -- (1.6,10.4) -- cycle;
		\draw [->-=0.5] (2.4,8.) -- (2.4,8.8);
		\draw [->-=0.5] (2.4,8.8) -- (2.4,9.6);
		\draw [->-=0.5] (2.4,9.6) -- (2.4,10.4);
		\draw [->-=0.5] (2.4,8.8) -- (3.,8.8);
		\draw [->-=0.5] (2.4,9.6) -- (3.,9.6);
		\node [] at (3.,9.) {\scriptsize $#1$};
		\node [] at (3.,9.8) {\scriptsize $#2$};
		\node [] at (2.,8.4) {\scriptsize $#3$};
		\node [] at (2.,9.2) {\scriptsize $#4$};
		\node [] at (2.,10.) {\scriptsize $#5$};
		\end{tikzpicture}
	}
	
	For the case of a disk, if a flux/rough boundary exist for a set of input data, then the $\overline{B}_p$ terms reduce to $\overline{B}_{n,n+1}$ where the half plaquette $\overline{B}_p$ lies between the open links $a_n$ and $a_{n+1}$. Specifically, the matrix elements are given by
	\begin{equation}
	\begin{split}
	& \bbra{\BpBasisEff{a'_n}{a'_{n+1}}{l_{n-1}}{l'_n}{l_{n+1}}}\overline{B}^t_{(n,n+1)}\bket{\BpBasisEff{a_n}{a_{n+1}}{l_{n-1}}{l_n}{l_{n+1}}}
	\\
	= &
	u_{a_n}u_{a_{n+1}}u_{a'_n}u_{a'_{n+1}}
	v_{l_n}v_{l'_n} f_{a_{n+1}t^{a'_{n+1}}}f_{ta_n{a'_{n}}}\\
	& ~~~~
	\times G^{l^*_{n+1}l_na^*_{n+1}}_{ta^{\prime*}_{n+1}l'_n}
	G^{l^*_nl_{n-1}a^*_n}_{a^{\prime*}_nt^*l'_n}.\\
	\end{split}
	\end{equation}
	The ground state is $
	GSD_{D^2}^{\text{flux}}=\text{tr}~(B_p \prod_n \overline{B}_{n,n+1} ) =1.$ 
	
\section{The correspondence for a disk}\label{sec:disk}
\subsection{Boundary spectrum on a disk}\label{subsec:BdryDisk}
The ground state is always non-degenerate on a disk. Throughout the paper we assume the parameter $\epsilon$ in equation \eqref{eq:FullHam} is small, so that the bulk is always in its ground state. We will comment on the relaxation of this assumption in the discussion section. The $n$-th excited state of the full system then corresponds to $n$ total boundary excitations created from the ground state. If there are $n_\alpha$ boundary excitations of type $\alpha$, $n_\beta$ of type $\beta$ etc., the degeneracy for this excited state then includes the number of possible distributions of these boundary excitations on these sites:
\begin{equation}
\binom{L}{n_\alpha}\binom{L-n_\alpha}{n_\beta}\binom{L-n_\alpha-n_\beta}{n_\gamma}\cdots\binom{n_\zeta+n_0}{n_\zeta},\nonumber
\end{equation}
where $n_0$ is the number of boundary sites that are not occupied by excitations. The set $\{n_\alpha\}$ satisfies $n=n_\alpha + n_\beta + \dots,~~n\leq L.$ Since these excitations are created from the ground state, they should conversely fuse into vacuum. Labeling the number of fusion channels as $g_n$, the degeneracy of such distribution of excitations $\{n_\alpha\}$ is
\begin{equation}
\text{deg}_{n,\{n_\alpha\}}(D^2)=  \frac{L!}{n_0!n_\alpha!n_\beta!n_\gamma!\cdots n_\zeta!}~g_n.
\label{eq:deg-n}
\end{equation}

Specifically for a charge/smooth boundary, the boundary excitations are all charges, while for a flux/rough boundary (if exists), they are all fluxes. 
	
{In the grand canonical ensemble, one can define the boundary spectrum with respect to a distribution of boundary excitations: for each set of $\{n_\alpha\}$, there is a a diagonal matrix 
\begin{equation}
\rho_{n,\{n_\alpha\}}=(\prod_{\alpha} d_\alpha^{~n_\alpha}) \mathbbm{1},
\label{eq:Qspectrum}
\end{equation}
where $\mathbbm{1}_{n,\{n_\alpha\}}$ is an identity matrix with dimensions deg$_{n,\{n_\alpha\}}\times$deg$_{n,\{n_\alpha\}}$. As introduced in \eqref{eq:DM}, this is the density matrix at the infinite-temperature and the constant-fugacity $z_\alpha=d_\alpha$ limit. The infinite temperature expression is possible to match with the entanglement computation in the bulk because it allows all the microstates on the boundary to occur with the equal probability. We will see in the next section that in the bulk entanglement computation, the reduced density matrix is a projection operator, such that all probable pure states have equal weights. The spectrum is sensitive to not only the total energy $n$, but the numbers of all types of boundary excitations.}

{The identification $z_\alpha=d_\alpha$ can be intuitively understood in the following way. Consider a mixture of gases with different types of particles labeled by $\alpha, \beta, \cdots$. The fugacity $z_\alpha$ then describes the tendency for a gas of type $\alpha$ to escape or expand, which is proportional to its pressure. 
Furthermore, the pressure is proportional to the number of the corresponding particles per volume. At high temperatures, all microstates appear with equal probability, so the number of particles $n_\alpha$ is proportional to the number of linearly independent microstates for each particle of type $\alpha$. The latter is asymptotically $d_\alpha$ at large $n_\alpha$ \cite{KPTEE}.}


\subsection{Entanglement spectrum for a disk subsystem}\label{subsec:EntDisk}
	
Now we turn to the reduced density matrix obtained from putting a string-net ground state on $S^2$, bipartitioning it into two disks $A, B$, and tracing out one of them\footnote{For general topological systems with finite correlation length, we require the both subregions to be much larger compared to the correlation length.}.  For convenience, we specify the cut to be of rough type, namely, the cut intersects links instead of passing through vertices. The rough cut and the smooth cut are equivalent in the computation of reduced density matrix because for a rough cut, the open links in $B$ are not free degrees of freedom: they must match with those in $A$. Doing the partial trace in A automatically leads to a partial trace in the open links of B, which is the same as doing a smooth cut in the first place.
\begin{figure}[htbp]
	\centering
	\includegraphics[]{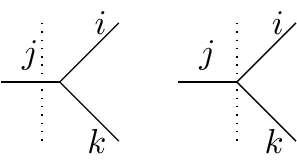}
	\caption{Types of entanglement cuts. Left: rough cut. The cut (dotted line) intersects some link $j$. Right: smooth cut. There is no intersection. The names ``rough'' and ``smooth'' are chosen to be the same as the ``rough'' and ``smooth'' boundary conditions introduced section \ref{sec:SN} because of their similar shapes.}
	\label{fig:rough}
\end{figure}

\begin{figure}[htbp]
	\centering
	\includegraphics[]{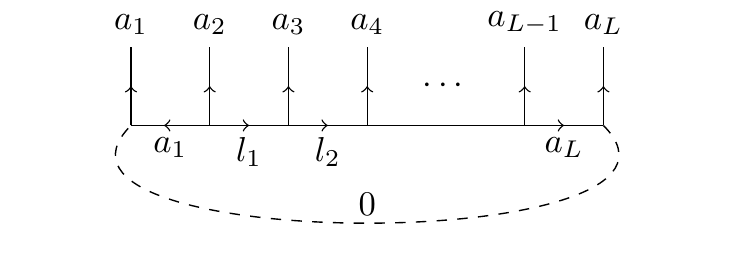}
	\caption{Basic tree-like configuration in a disk-shaped region $A$.}
	\label{fig:disk}
\end{figure}

The remaining disk $A$ can be smoothly deformed into the following configuration  \ref{fig:disk}. 

The resultant reduced density matrix was calculated in \cite{WenTEE} and is diagonal,
\begin{equation}
	\label{eq:WenRDM}
	\langle \left\{ a', l'\right\}| \rho_A|\left\{a, l\right\}\rangle = \delta_{\{a\},\{a'\}}\delta_{\{l\},\{l'\}}D^{1-L}\prod_{m=1}^{L} {d_{a_m}}{}.
\end{equation}
	
To make a connection with the boundary quasiparticle spectrum discussed in the last section \ref{subsec:BdryDisk}, one can recast the above formula \eqref{eq:WenRDM} using fiber fusion category language developed in \cite{Luo}, as 
	\begin{equation}
	\label{eq:diskRDM}
	\rho_A=D P_0 (\alpha^{\otimes L}),
	\end{equation}
	where $\alpha$ is a diagonal matrix with rank $|I|$, the number of string types in the input category. The entries are $\alpha_j=d_j/D$. We define a product ``$\otimes$'' for $\alpha$ as
	\begin{equation}
	\alpha_k^{\otimes 2} = (\alpha \otimes \alpha)_k=\oplus_{i,j\in I}~ \alpha_i\alpha_j\delta_{jik^*}.
	\end{equation}
	The resultant $\alpha^{\otimes 2}$ is again a diagonal matrix, and generally one has
	\begin{equation}
	\label{eq:tensor}
	\alpha_k^{\otimes L}=(\alpha^{\otimes (L-1)}\otimes \alpha)_k=\oplus_{i,j\in I} ~\alpha_i^{\otimes (L-1)} \alpha_j \delta_{ijk}^*.
	\end{equation}
	Operator $P_j$ projects onto the $j$-component of the $\alpha^{\otimes L}$ matrix, so that $P_0$ implements the constraint that all open links should fuse to vacuum.
	
Suppose the nontrivial open links with label $a\neq 0$ appears $n_a$ times in the configuration $\{a_1,a_2,\cdots,a_L\}$ of \ref{fig:disk}, so that the total number of nontrivial open links is $n=\sum_{a\neq 0} n_a$. The diagonal reduced density matrix $\rho_A$ in \eqref{eq:diskRDM} then consists of blocks of smaller diagonal matrices 
\begin{equation}
\rho_A=\oplus_n~\rho_{(A,n)}=\oplus_n \oplus_{\{a_n\}}~\rho_{A;n,\{n_a\}},
\end{equation}
with the direct sum over $\{a_n\}$ subject to the constraint $n=\sum_{a\neq 0} n_a$. {Then the dimension of each $\rho_{(A,n)}$ is exactly equal to the degeneracy of the $n$-th excited state in the boundary spectrum of a disk. Actually there is a more refined match: the dimension of $\rho_{A;n,\{n_a\}}$ equals deg$_{n,\{n_a\}}$ in \eqref{eq:deg-n}. Furthermore, the value of each entry related to the distribution $\{n_\alpha\}$ are the same for the density matrix on the boundary \eqref{eq:Qspectrum} and the reduced density matrix up to an overall factor,
\begin{equation}
\rho_{A;n,\{n_a\}}=D^{1-L} \rho_{n,\{n_a\}}.
\end{equation}
This leads to the correspondence between the bulk entanglement spectrum and the boundary excitation distribution spectrum.
}

	In the above discussions, we have taken $\alpha$ to be a diagonal matrix with rank $|I|$, which implies that the corresponding boundary theory is of charge/smooth type. Namely, the boundary excitations $n_\alpha, n_\beta, \cdots, n_\zeta$ are all charges. More generally for a boundary theory with Frobenius algebra $I_\mathcal{A}\subsetneq I$, it is tempting to take $\alpha$ to be of rank $|I_\mathcal{A}|$, where each $\alpha_j=d_j/d_\mathcal{A}$ has $j\in I_\mathcal{A}$. However at this moment it is not clear what the physical meaning is, to constrain the open links intersecting an entanglement cut to $I_\mathcal{A}$.
	
	\subsection{Examples}\label{subsec:DiskExample}
	To be concrete, we discuss the two familiar examples of toric code model and the doubled Fibonacci model.
	
	For the toric code, the input data form the representation category of the $\mathbb{Z}_2$ group. The label set $I=\{0,1\}$, $0*=0$ and $1^*=1$. The quantum dimensions are $d_0=d_1=1$, the nonzero fusion rules are $\delta_{000}=\delta_{011}=\delta_{101}=\delta_{110}=1,$
	with the $G$ symbols being
	\begin{equation}
	G_{kln}^{ijm}=\delta_{ijm}\delta_{klm^*}\delta_{jkn^*}\delta_{inl}.
	\end{equation}
	String-net model outputs four types anyons: $\{1,e,m,\epsilon=e\oplus m\}$, where $e$ is a $\mathbb{Z}_2$ charge and $m$ a $\mathbb{Z}_2$ flux.
	
	There are two Frobenius algebras, i.e. boundary conditions for this input data. One is the trivial $\mathcal{A}_0=0$, which defines a charge boundary condition. Excitations on the charge boundary are identified with $0$ and $1$ or equivalently $\unit$ and $e$. The other Frobenius algebra is $\mathcal{A}_1=0\oplus 1$, with $I_\mathcal{A}=I$. This is a flux boundary condition. Boundary quasiparticles are identified with $\unit$ and $m$. For both types of boundaries, there is only one type of nontrivial excitation, 
	\begin{equation}
	\text{deg}_n(\mathbb{Z}_2; D^2)=\binom{L}{n} g_n,~~g_n(\mathbb{Z}_2; D^2)=1.
	\end{equation}
	This is typical for models with abelian fusion rules.
	
	The simplest non-abelian example is the doubled Fibonacci model, where $I=\{0,2\}$ (sometimes also denoted  as $\{1,\tau\}$) with $0^*=0$, $2^*=2$. Let $\phi=(1+\sqrt{5})/2$ be the golden ratio, then the quantum dimensions are given by $d_0=1$ and $d_2=\phi$. The nonzero fusion rules are $\delta_{000}=\delta_{022}=\delta_{202}=\delta_{220}=\delta_{222}=1,$
	and the independent $G$ symbols are 
	\begin{equation}
	\begin{split}
	& G_{000}^{000}=1,~G_{022}^{022}=G_{222}^{022}=1/\phi, \\
    & G_{222}^{000}=1/\sqrt{\phi},~G_{222}^{222}=-1/\phi^2.
	\end{split}
	\end{equation}
	The bulk quasiparticles are labeled by $\{0\bar{0},0\bar{2},2\bar{0},2\bar{2}\}$, or sometimes $\{1\bar{1}, 1\bar{\tau}, \tau\bar{1}, \tau\bar{\tau}\}$. The above input category gives rise to two Frobenius algebras: $\mathcal{A}_0=0$, which defines a charge boundary condition and $\mathcal{A}_1=0\oplus 2$, giving a flux boundary condition. The latter leads to a nontrivial multiplication $f_{222}=\phi^{-3/4}$. These two Frobenius algebras are Morita equivalent, i.e. there is a map between all irreducible $A_0$ modules and all irreducible $A_1$ modules which preserves the fusion rules. Being Morita equivalent to each other means that the two Frobenius algebras give rise to the same boundary condition \cite{Bdry1}.
	
	The degeneracies are again characterized by $\text{deg}_n(\text{dFib},D^2)=\binom{L}{n} g_n(\text{dFib},D^2)$, with
	\begin{equation}
	g_n(\text{dFib};D^2) = F_{n-1}.
	\end{equation}
	Here $F_n$ the Fibonacci sequence satisfying $F_1=F_2=1$, $F_n=F_{n-1}+F_{n-2}$ for $n>2$.
	
	\section{The correspondence for a cylinder}\label{sec:cyl}
	
	\subsection{Boundary theory on a cylinder}\label{subsec:BdryCyl}
\begin{figure}[h]
		\centering
		\includegraphics[]{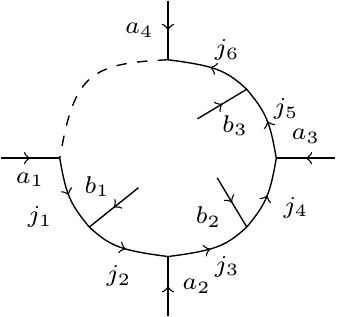}
		\caption{Effective configuration of on a cylinder.}
		\label{fig:cyl}
	\end{figure}

	Different boundary conditions or Frobenius algebras $\mathcal{A}, \mathcal{B}$ can be chosen for the two boundaries $A$ and $B$ of a cylinder. In the case without bulk excitations, one can deform the bulk graph by (dual) Pachner moves so that the cylinder graph shrinks to a ring with open links on both sides of the ring, see for example figure \ref{fig:cyl}
	
	Both the ground state degeneracy (GSD) and the topological quasiparticles are classified by $\mathcal{A}-\mathcal{B}$ bimodules. We refer to section 6 of \cite{Bdry1} for its detailed mathematical structure.
	
	If both boundaries are simply charge boundaries $\mathcal{A}=\mathcal{B}=\mathcal{A}_0$, then the $\mathcal{A}_0-\mathcal{A}_0$ bimodule is the entire label set $I$, so that both the GSD and the quasiparticles on the cylinder are labeled by the string types $j\in I$. Denote the total number of sites as $L=L_A+L_B$, and again suppose the charge excitations of type $\{\alpha\}$ have number $\{n_\alpha\}$, then the degeneracy for such distribution of excitations will be of a familiar form \eqref{eq:deg-n}, for there is no need to distinguish the two boundaries. But the factor $g_n$ should now take into account the degenerate ground state subspace on a cylinder. For abelian models this simply amounts to $g_n=\text{GSD}=|I|$. For non-abelian cases, the multiple fusion channels and degenerate ground states combine in a nontrivial way.  Other kinds of boundaries including the flux type need to be analyzed case by case. The spectrum becomes more complicated with general combinations of the two boundary conditions. 
	
{We can again define the boundary excitation distrubution spectrum for a cylinder. For a distribution $\{n_\alpha, n_\beta\}$ on the two boundaries, the density matrix in the double limits is
\begin{equation}
\rho_{n,\{n_\alpha\}}=(\prod_{\alpha}z_\alpha^{n_\alpha}\prod_{\beta}z_\beta^{n_\beta})~\mathbbm{1},
\end{equation}
with $\mathbbm{1}$ the identity matrix of dimensions deg$_{n, \{n_\alpha, n_\beta\}}\times$ deg$_{n, \{n_\alpha, n_\beta\}}$, and $z_\alpha=d_\alpha,$ $z_\beta=d_\beta$. Only in the special cases where the two boundary conditions are the same, can we combine the two products. Generally the meaning of boundary excitations for the two boundaries can be different.
} 

For the toric code model, there are three possible boundary conditions for the toric code model on a cylinder, labeled by the different Frobenius algebras: (i) $\mathcal{A}=\mathcal{B}=0$, both being charge boundaries; (ii) $\mathcal{A}=0,$ $\mathcal{B}=0\oplus 1$ or $\mathcal{B}=0,$ $\mathcal{A}=0\oplus 1$, the mixed boundaries; and (iii)  $\mathcal{A}=\mathcal{B}=0\oplus 1$, both being flux boundaries. Cases (i)(iii) both give twofold GSD, so that
\begin{equation}
\begin{split}
& g_n(\mathbb{Z}_2; \text{cylinder, charge})=g_n(\mathbb{Z}_2; \text{cylinder, flux})\\
= & \begin{cases}2, & 0\leq n\leq L~\text{and}~ n\in 2\mathbb{Z}\\
0, & \text{else}\end{cases}
\end{split}
\label{eq:Z2ChargeGSD}
\end{equation}
where the $n\in 2\mathbb{Z}$ constraint arises from the pair creation of $\mathbb{Z}_2$ excitations and $L=L_A+L_B.$ Degeneracy of the $n$-th excited state is given by the usual 
\begin{equation}
\text{deg}_n(\mathbb{Z}_2; \text{cylinder, charge/flux})=g_n\binom{L}{n}.
\label{eq:Z2ChargeDeg}
\end{equation}
So in these two cases, the counting behaves as if there is only one boundary. By comparison, case (ii) leads to a non-degenerate ground state subspace and
\begin{equation}
g_n(\mathbb{Z}_2; \text{cylinder, mixed})=\begin{cases} 1, & 0\leq n<L\\
0, & \text{else}.\end{cases}
\label{eq:Z2Mixed}
\end{equation}
Now we need to distinguish the excitations on different boundareis. The degeneracy of the $n$-th excited state with distribution $\{n_\alpha, n_\beta\}$ is 
\begin{equation}
\text{deg}_{n, \{n_\alpha,n_\beta\}}(\mathbb{Z}_2; \text{cylinder, mixed})=g_n \binom{L_A}{n_A}\binom{L_B}{n_B}.
\end{equation}
Summation of the above degeneracy over all possible distributions $\{n_\alpha,n_\beta\}$ satisfying $n_\alpha+n_\beta=n$ gives the usual degeneracy of the $n$-th excited state on a cylinder deg$_n(\mathbb{Z}_2; \text{cylinder, mixed})$.
Intuitively, the above three cases can be understood in terms of the anyon condensation language \cite{Bais1, Bais2, Bombin, Bais3, Kapustin, BombinNested, Levin, Barkeshli, Kong, Hung, Lan, Bernevig2}. Case (i) corresponds to the fluxes $m$ condensing on both boundaries and cannot be distinguished with the vacuum $\unit$, while the anyons with nontrivial charges $e$ and $\epsilon$ becomes excitations on the boundary. So the two GSD correspond to $\unit$ and $m$. For case (iii) the GSD is two again, but is now labeled by $\unit$ and $e$ and correspond to the condensation of $e$ particles. Then case (ii) is that of the mixed boundaries. The corresponding GSD is only one, since the $m$ flux can be distinguished from vacuum $1$ on the flux boundary, while the $e$ charge can be distinguished from $1$ on the charge boundary. So all four degenerate ground states on the torus can be distinguished on the boundaries. There is no $n\in 2\mathbb{Z}$ constraint in $g_n$ in this case, because one can for example create a pair of charges $e$, move one of them to boundary $A$ and the other to $B$. On one of these two boundaries, $e$ is identified with vacuum and thus gives no excitation energy. (We note that the above intuitive understanding is helpful but not rigorous; the general relationship between the Frobenius algebra formalism and the anyon condensation picture of boundary theories is yet to be derived.)

In a similar fashion, for $\mathbb{Z}_N$ models we also have cases (i) of two charge boundaries and (iii) of two flux boundaries both with $g_n=N$, and case (ii) of mixed charge and flux boundaries with $g_n=1$. Additional boundary types other than the charge and flux ones are also possible, giving rise to more combinations.
	
{For the simplest non-abelian case of doubled Fibonacci, there is only one type of boundary condition, as  reviewed in section \ref{subsec:DiskExample}. Furthermore, there is only one type of nontrivial boundary excitation. So the generalized density matrix contains only one subscript,
\begin{equation}
\rho_n=\phi^n \mathbbm{1}.
\label{eq:FibQ}
\end{equation}
Above $\mathbbm{1}$ is an identity matrix of dimensions deg$_n\times$deg$_n$.} The degeneracy is
\begin{equation}
\text{deg}_n(\text{dFib; cylinder, charge})=L_n \binom{L}{n}, 
\label{eq:FibCyl}
\end{equation}
with $L_n$ the Lucas sequence $2, 1, 3, 4, 7, 11, \cdots$. The $n=0$ case gives the twofold ground state degeneracy.

\subsection{Entanglement spectrum on a cylindrical subsystem}\label{subsec:EntCyl}
In this section, we discuss the entanglement properties of the string-net model on a torus, where subsystem $A$ will have the topology of a cylinder. The focus will be on the abelian $\mathbb{Z}_N$ models, especially the toric code $\mathbb{Z}_2$ case, and leave the non-abelian example to the next section.
	
\begin{figure}[htbp]
		\centering
		\includegraphics[]{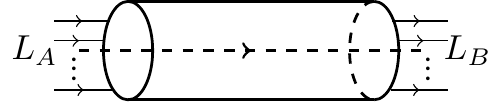}
		\caption{The entanglement cut generates $L_A$ and $L_B$ open links on the two boundaries of the cylinder.}
		\label{fig:CylCut}
	\end{figure}

	For toric code model in the quasiparticle basis, a general ground state can be written as $\ket{\Psi}=c_1 \ket{\unit}+c_2 \ket{m}+c_3 \ket{e}+c_4 \ket{\epsilon}$. Suppose the entanglement cut intersects $L_A+L_B$ links as in fig. \ref{fig:CylCut}, the diagonalized reduced matrix then consists of four blocks
	\begin{equation}
	\rho_A=2^{-L_A+1-L_B+1} \left(\begin{matrix} |c_1|^2 \mathbbm{1} & & & \\ & |c_2|^2 \mathbbm{1} & & \\ & & |c_3|^2 \mathbbm{1} & \\ & & & |c_4|^2 \mathbbm{1} \end{matrix}\right),
	\end{equation}
	where each $\mathbbm{1}$ is an identity matrix of dimensions $2^{L_A-1+L_B-1}\times 2^{L_A-1+L_B-1}$. The number $2^{L_A-1+L_B-1}$ is the summation of the degeneracies in the boundary excitation spectrum on a cylinder from the last section:
	\begin{equation}
	2^{L_A+L_B}=\sum_{n=0}^{L_A+L_B} deg_n(\mathbb{Z}_2; \text{cylinder, mixed}).
	\label{eq:CylMixed}
	\end{equation}
(Since the model is abelian with all quantum dimensions $d_\alpha=1$, {the fugacities are trivial}, and the boundary excitation distribution spectrum is flat.)

On the entanglement side, deg$_n$ is understood as the number of configurations in fig. \ref{fig:cyl} with altogether $n$ nontrivial open links that intersect the entanglement cut. The difference factor of $4$ between $2^{L_A+L_B}$ and $2^{L_A-1+L_B-1}$ gives the topological entanglement entropy, which can be easily read out from the reduced density matrix as
	\begin{equation}
	\begin{split}
	S(\mathbb{Z}_2; \text{cylinder})= & (L_A+L_B)\log 2-2\log 2\\
	& -\sum_{\J} |c_\J|^2 \log |c_\J|^2.
	\end{split}
	\end{equation}
	The first term is the usual area law, while second term arises from the topological entanglement entropy due to the two boundaries of the cylinder as already observed in \cite{MES}, and the third term is the Shannon entropy from the combination of different $\J$'s. 
	For $\mathbb{Z}_N$ models, this is similarly
	\begin{equation}
	\begin{split}
	S(\mathbb{Z}_N; \text{cylinder})=& (L_A+L_B)\log N-2\log N\\
	& -\sum_{\J} |c_\J|^2 \log |c_\J|^2, 
	\end{split}
	\end{equation}
	with $\J$ running over the $N^2$ quasiparticles of thte $\mathbb{Z}_N$ model.
	
\begin{figure}[htbp]
		\centering
		\includegraphics[]{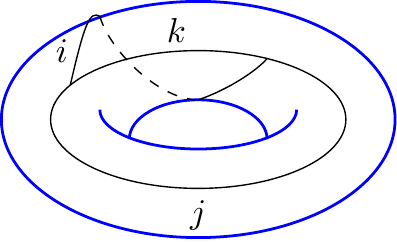}
		\caption{The simplest graph on torus contains two vertices, three links and one plaquette. Links $i$ and $k$ winds the meridian of the torus, while $j$ and $k$ winds the longitude. }
		\label{fig:SimpleGraph}
	\end{figure}

	From equation \eqref{eq:CylMixed}, we have a correspondence between the cylinder boundary excitation spectrum with mixed boundary condition \eqref{eq:Z2Mixed} and the entanglement spectrum. A natural question, then, would be to understand how the other two types of boundaries (i)(iii) can be realized from the entanglement side. To this end, we introduce the simplest graph on a torus given by Fig.\ref{fig:SimpleGraph}, with three links, two vertices and one plaquette. All other more complicated graphs can be obtained from this simplest graph through the (dual) Pachner moves introduced in the first section.
	
\begin{figure}[htbp]
		\centering
	\includegraphics[scale=0.25]{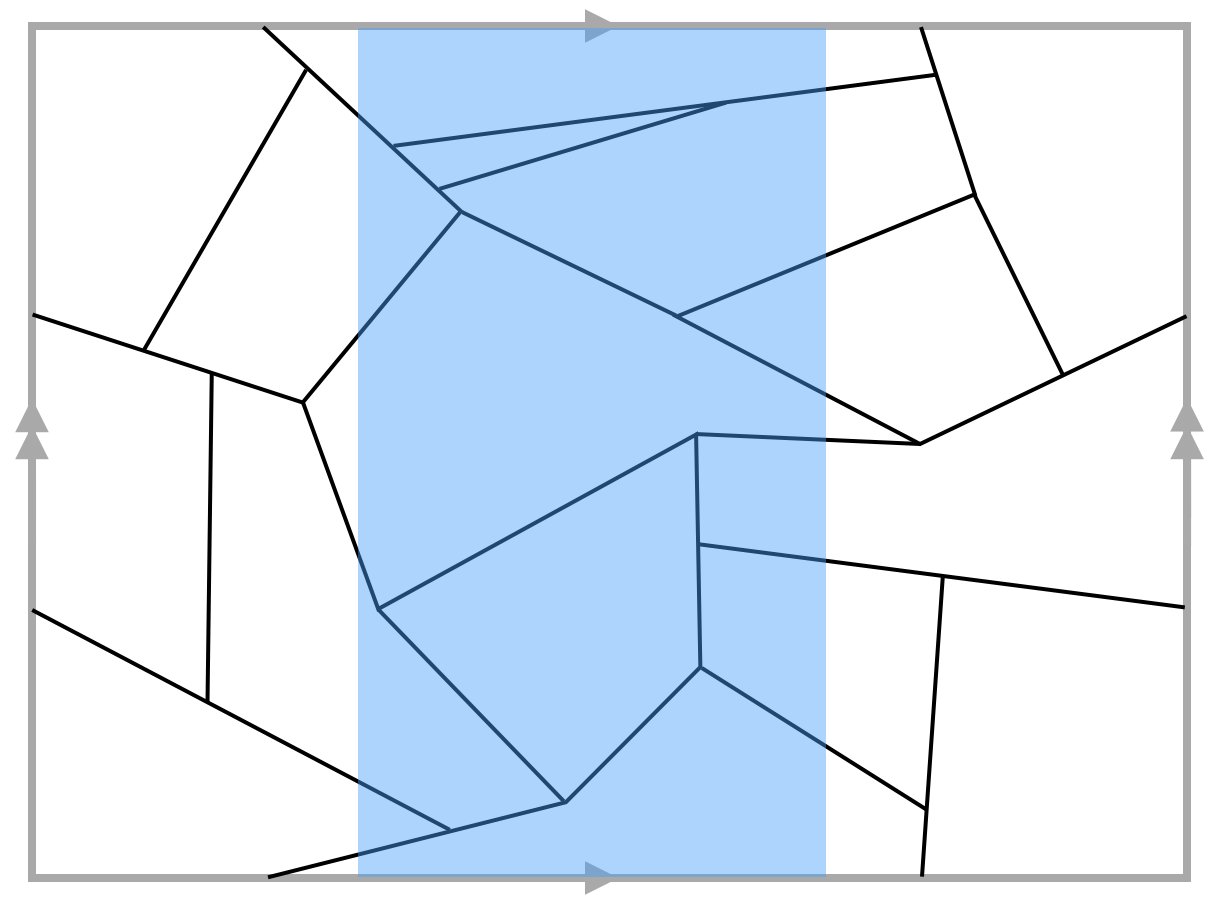}
		\caption{Example of the ``disk+cylinder'' bipartition beyond the simplest graph. The torus is constructed from a rectangle by gluing both pairs of opposite edges. The blue shaded region is the subsystem where the graph does not see any non-contractible loop, while the unshaded region sees one non-contractible loop.}
		\label{fig:shade}
	\end{figure}

	A general ground state on the simplest graph can be written as a superposition of different configurations $\ket{ikj}$ . For our toric code example, the relevant states are \cite{Thesis}
	\begin{equation}
	\begin{split}
	& \ket{\unit}=\frac{1}{\sqrt{2}} (\ket{000}+\ket{110}),
	\ket{m}=\frac{1}{\sqrt{2}} (\ket{000}-\ket{110}),\\
	& \ket{e}=\frac{1}{\sqrt{2}} (\ket{011}+\ket{101}),
	\ket{\epsilon}=\frac{1}{\sqrt{2}} (\ket{011}-\ket{101}).\\
	\end{split}
	\label{eq:Z2MES}
	\end{equation}
	If the entanglement cut splits the non-contractible loop labeled by $j$, for example let $i, k$ belong to subsystem $A$ and $j$ belong to subsystem $B$, then $A$ is topologically a ``cylinder'' while $B$ is a ``disk'', for there is no non-contractible loop in the graph of $B$. Then each of these four states \eqref{eq:Z2MES} gives a trivial reduced density matrix with zero entanglement entropy by itself, due to the small number of total links in this graph. To obtain more general results, we can do dual Pachner moves to complicate the graph, but comply with one important constraint: we want subsystems $A$ and $B$ to stay as a cylinder and a disk, respectively. One example is shown in the fig. \ref{fig:shade}.

\begin{figure}[htbp]
		\centering
	\includegraphics[]{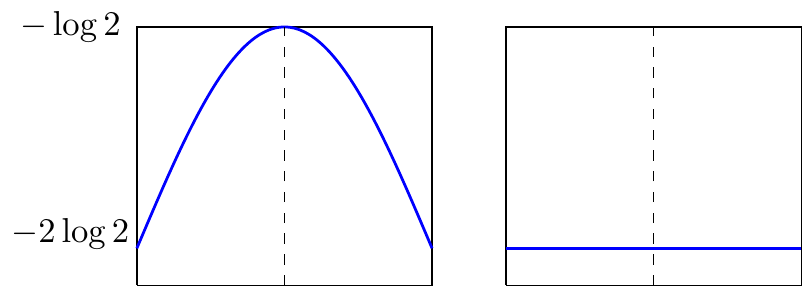}
		\caption{(Color online) The entanglement entropy is only sensitive to topological entanglement sectors. The horizontal direction denotes the ratio $|c_{\J_1}|^2/|c_{\J_2}|^2$, while the vertical direction is the negative of the topological entanglement entropy. \textbf{Left}: superposing ground states in different minimally entangled sectors. $\J_1$ and $\J_2$ belong to different sectors (for example one can take $\J_1=\unit, \J_2=e$). \textbf{Right}: superposing ground states in the same minimally entangled sector. $\J_1$ and $\J_2$ belong to the same sector (for example one can take $\J_1=e, \J_2=\epsilon$).}
		\label{fig:MES-Z2}
	\end{figure}

	Generally the reduced density matrix is  \cite{Luo}
	\begin{equation}
	\rho_A=2^{L-1}\left[(|c_1|^2+|c_2|^2)\mathbbm{1}\oplus(|c_3|^2+|c_4|^2)\mathbbm{1}\right],\nonumber
	\end{equation}
	with $\mathbbm{1}$ a $2^{L-1}\times 2^{L-1}$-dimensional identity matrix and $L$ is the total number of links intersecting the entanglement cut. The entanglement entropy becomes
	\begin{equation}
	\label{eq:S-Z2}
	\begin{split}
	S (\mathbb{Z}_2, & \text{cylinder+disk})= L \log 2-\log 2\\
	& -\sum_j (\sum_{\J} |c_\J|^2 M_{\J j}) \log (\sum_{\J} |c_\J|^2 M_{\J j}),
	\end{split}
	\end{equation}
	where the decomposition matrix $M$ is $4\times 2$-dimensional, with the first subscript taking values from $\{\unit,m, e, \epsilon\}$, i.e. the output category, and the second subscript from $\{0,1\}$, the input category.\footnote{The decomposition $M$ similar to but different from the tunneling matrix introduced in \cite{Lan} in the range and meaning of the second subscript. For details see \cite{Luo}.} Specifically, 
	\begin{equation}
	M_{\unit 0}=M_{m0}=1,~~M_{e1}=M_{\epsilon 1}=1,~~\text{else}=0.
	\end{equation}
	The rows of $\unit, m$ are exactly the same, so that states $\ket{1}$ and $\ket{m}$ are not distinguishable from the perspective of reduced density matrix. Similar phenomenon happens for the states $\ket{e}$ and $\ket{\epsilon}$: whatever relative weights we set for these two states, the reduced density matrix and the entanglement entropy are always the same. For this reason, three of the authors generalized the concept of minimally entangled states in \cite{MES} to that of the \textit{minimally entangled sectors} \cite{Luo}. States that cannot be distinguished from the entanglement perspective are understood to be in the same sector. In our example, there are two sectors $\{\ket{\unit}, \ket{m}\}$ and $\{\ket{e},\ket{\epsilon}\}$, which is illustrated in fig. \ref{fig:MES-Z2}. It is possible to reach the maximum entanglement entropy or the minimum topological entanglement entropy only when superposing two ground states $\J_1$ and $\J_2$ that come from different sectors.
	
	For $\Z_N$ models beyond $N=2$, the phenomenon is similar. In the quasiparticle basis, there are $N^2$ elementary ground states labeled by 
	$\J=(g,j)$, with $g,j \in \{0,1,\cdots,N-1\}$ denoting fluxes and charges, respectively. 
	\begin{equation}
	\label{eq:ZNstates}
	\ket{(g,j)}=\frac{1}{\sqrt{N}}\sum_{i,k}e^{2\pi \mathbf{i} ig/N}\delta_{jik^*} \ket{ikj}.
	\end{equation}
	One observes that the flux and charge degrees of freedom are assigned to the two non-contractible loops $i$ and $j$ of the torus (fig. \ref{fig:SimpleGraph}), respectively \footnote{We have used the bold $\mathbf{i}$ to denote the imaginary unit, in order to distinguish it from the link $i$. The phase factor $e^{2\pi \mathbf{i} ig/N}$ is the half braiding tensor in the sense of \cite{StringNet, Full}. For the $\Z_N$ case it translates the flux degree of freedom $g$ to link degree of freedom $i$.}. These states are grouped into $N$ sectors with the decomposition matrix $M_{(g,\mu),j}=\delta_{\mu j}.$  Namely, those states that are labeled by the quasiparticles with different fluxes but same charge numbers are in the same sector. In other words, the fluxes are undetectable using entanglement entropy. This is analogous to the situation in the boundary excitation spectrum with two charge boundaries (case i) of toric code discussed in the last section: there the fluxes were also ``condensed'' and thus undetectable in the energy spectrum. The ground state degeneracy $N$ on a genuine cylinder is matched with the order of the minimally entangled sector that contains $\ket{\unit}$, while deg$_n$ on a genuine cylinder is matched to the dimension of the block in the diagonal reduced density matrix that contains $n$ nontrivial open links. We note that minimally entangled sectors is characteristic of non-chiral topological ordered systems, and is not present in chiral cases \cite{Fradkin}.
	
	Back to the toric code example, the ground state degeneracy \eqref{eq:Z2ChargeGSD} is matched with the order of the minimally entangled sector $\{\ket{\unit}, \ket{m}\}$, while the excited state degeneracy deg$_n$ in \eqref{eq:Z2ChargeDeg} is matched with the dimension of the specific block in the reduced density matrix which contains configurations with $n$ nontrivial open links.
	
	In the above we split $j$-loop in the graphs using the entanglement cut and find the correspondence between bulk entanglement and the boundary energy spectra of a cylinder with two charge boundaries (case i). Alternatively, we can split the non-contractible loop labeled by $i$. One reads from equation \eqref{eq:Z2MES} that for the simplest graph, the minimally entangled sectors now change to $\{\ket{\unit},\ket{e}\}$ and $\{\ket{m}, \ket{\epsilon}\}$. As long as one keeps the subsystem $A$ and $B$ to be cylinder and disk topologically, one can generate more complicated graphs and obtain the same sectors. This new set of sectors appear as if the charges are invisible to the entanglement spectrum, similar to the ``condensation'' of charges in the boundary condition (iii) for a genuine cylinder. The ground state degeneracy of the genuine cylinder again matches with the order of the minimally entangled sector which contains $\unit$, i.e., GSD$=|\{\ket{\unit},\ket{e}\}|=2$. The story goes in parallel for $\mathbb{Z}_N$ models. Now the $N^2$ states are regrouped into $N$ sectors, and those states whose corresponding quasiparticles have same flux but different charges will be in the same sector. \footnote{If the cut splits the $i$-loop, entanglement entropy will have an additional term of $\log D=\log N$ due to the Fourier transform between flux degrees of freedom $g$ and the string types $i$ in \eqref{eq:ZNstates}.}
	
	We summarize the toric code results in the following table \ref{tableZ2}.
	\begin{table}[htbp]
		\centering
		\begin{tabular}{ |c | c |}
			\hline
			\bf{Boundary} & \bf{Bulk Entanglement} \\ 
			\hline
			(i) charge boundaries & cylinder+disk cut, split $j$-loop  \\  
			\hline
			$\mathcal{A}=\mathcal{B}=0$ & $\{\unit,m\}, \{e, \epsilon\}$  \\
			GSD$=2$ & $|\{\unit,m\}|=2$ \\
			\hline
			(ii) mixed boundaries & cylinder+cylinder cut\\
			\hline
			$\mathcal{A}=0, \mathcal{B}=0\oplus 1$ & $\{\unit\}, \{e\}, \{m\}, \{\epsilon\}$ \\
			GSD$=1$ & $|\{\unit\}|=1$ \\
			\hline
			(iii) flux boundaries & cylinder+disk cut, , split $i$-loop\\
			\hline
			$\mathcal{A}=\mathcal{B}=0\oplus 1$ & $\{\unit,e\}, \{m, \epsilon\}$  \\
			GSD$=2$ & $|\{\unit,e\}|=2$ \\
			\hline
		\end{tabular}
		\caption{Summary of the $\mathbb{Z}_2$ model.The Frobenius algebras $\mathcal{A}, \mathcal{B}$ specify the boundary conditions of the two boundaries on a cylinder. On a torus there are two non-contractible loops labeled by $i$ (meridian) and $j$ (longitudinal). Varying the entanglement cuts changes structure of the reduced density matrix and gives rise to different minimally entangled sectors. (The ``cylinder+disk'' is a shorthand for taking the subsystems $A$ and $B$ to be cylinder and disk, respectively.) Cardinality of the minimally entangled sector that contains state $\unit$ matches with the ground state degeneracy on a cylinder with corresponding boundary conditions. Furthermore, the full entanglement spectrum matches with the boundary excitation distribution spectrum.}
		\label{tableZ2}
	\end{table}
	
\subsection{Non-abelian case}\label{subsec:Fib}
We now extend the discussion to non-abelian models. We start with the doubled Fibonacci case. Since there is only one type of boundary condition and one type of nontrivial boundary excitation, the boundary excitation distribution spectrum simply reduces to the boundary excitation spectrum. Hence, similarly to the $\mathbb{Z}_N$ case in \ref{subsec:EntCyl}, the boundary spectrum corresponds to doing entanglement cut by splitting the $j$-loop, and taking the subsystems $A$ and $B$ to be cylinder and disk, respectively.

On a simplest graph of a torus, the degenerate ground states are
\begin{equation}
\begin{split}
\ket{0\overline{0}}=&\frac{1}{1+\phi^2}\left(\ket{000}+\phi\ket{022}\right);\\
\ket{0\overline{2}}=&\frac{1}{1+\phi^2}\left(\ket{202}+e^{4\pi i/5}\ket{220}-\sqrt{\phi}e^{2\pi i/5}\ket{222}\right);\\
\ket{2\overline{0}}=&\frac{1}{1+\phi^2}\left(\ket{202}-e^{\pi i/5}\ket{220}+\sqrt{\phi}e^{3\pi i/5}\ket{222}\right);\\
\ket{2\overline{2}}=&\frac{1}{1+\phi^2}\left(\ket{000}-\phi^{-1}\ket{022}+\ket{202}+\ket{220}\right.\\
& \left.+\phi^{-3/2}\ket{222}\right).
\end{split}
\end{equation}

If the entanglement cut splits the $j$-loop,  the decomposition matrix is given by
\begin{equation}
M_{0\bar{0},0}=M_{2\bar{2},0}=1,~M_{0\bar{2},1}=M_{2\bar{0},1}=M_{2\bar{2},1}=1.
\end{equation}
There are three minimally entangled sectors in this case: $\mathcal{S}_1=\{0\bar{0}\}$, $\mathcal{S}_2=\{0\bar{2},2\bar{0}\}$ and $\mathcal{S}_3=\{2\bar{2}\}$. The first two sectors give the following entanglement entropies (where $L$ is the total number of links intersected by the entanglement cut):
\begin{equation}
S_A^{\mathcal{S}_1}=a L-\log  D,~
S_A^{\mathcal{S}_2}=a L-\log D+\log\phi.
\end{equation}
For the third sector, the reduced density matrix has the structure
\begin{equation}
\label{eq:FibRDM}
\rho_A^{\mathcal{S}_3}=\frac{1}{\phi^2}\rho_A^{\mathcal{S}_1}\oplus \frac{1}{\phi}\rho_A^{\mathcal{S}_2},
\end{equation}
so that we have
\begin{equation}
S_A^{\mathcal{S}_3}=\frac{1}{\phi^2} S_A^{\mathcal{S}_1} + \frac{1}{\phi}S_A^{\mathcal{S}_2}-\frac{1}{\phi^2}\log\frac{1}{\phi^2}-\frac{1}{\phi}\log\frac{1}{\phi}.
\end{equation}

For a general state $|\Psi\rangle = \sum_{\J} c_{\J} |\J\rangle$, the properties are the following. (i) Superposing states in the same sector does not change the entanglement entropy. (ii) For general superpositions among the three sectors, the maximal topological entanglement entropy is reached when $|c_{\mathcal{S}_1}|^2=1$ and $|c_{\mathcal{S}_2}|^2=|c_{\mathcal{S}_3}|^2=0$. Minimal topological entanglement entropy is reached when $|c_{\mathcal{S}_3}|^2=1$ and $|c_{\mathcal{S}_1}|^2=|c_{\mathcal{S}_2}|^2=0$. (iii) Specifically, superposing states in sectors $\mathcal{S}_1$ and $\mathcal{S}_2$, and topological entanglement entropy decreases linearly from $\log D$ to $\log D-\log \phi$ as $|c_{\mathcal{S}_2}|^2/|c_{\mathcal{S}_1}|^2$ goes from zero to infinity. (iv) Superposing sectors $\mathcal{S}_1$ and $\mathcal{S}_3$, the topological entanglement entropy again decrease monotonically, from $\log D$ to $\log D-(\frac{D}{\phi^2}+\frac{1}{\phi})\log\phi$ as the weight of $\mathcal{S}_3$ increases. (v) Similarly, superposing  $\mathcal{S}_2$ and $\mathcal{S}_3$ leads to monotonically decreasing topological entanglement entropy as the weight of $\mathcal{S}_3$ increases. 

The entanglement spectrum again shares the same levels and degeneracies with the boundary excitation (distribution) spectrum \eqref{eq:FibQ} {at infinite temperature and fixed fugacities}. In this correspondence, the boundary GSD is no longer simply matched with the order of the minimally entangled sector containing $\ket{\unit}=\ket{0\bar{0}}$. A more precise criteria is that \textit{GSD on the boundary = the number of $\J$ that has nontrivial entry $M_{\J 0}\neq 0$ in the decomposition matrix that appears in the bulk entanglement spectrum.} It reduces to the order of minimally entangled sectors containing $\ket{\unit}=\ket{0\bar{0}}$ for the $\mathbb{Z}_N$ case. In the doubled Fibonacci example, the two relevant states are $\ket{0\bar{0}}$ and $\ket{2\bar{2}}$. For $n\geq 1$, we rewrite $L_n$ in terms of the Fibonacci sequence $L_n=2F_{n-1}+F_n$, and notice that the first $F_{n-1}$ term corresponds to the number of channels for fusing $n$ nontrivial charges to vacuum $\ket{0\bar{0}}$, while the remaining $F_{n-1}+F_n$ corresponds the number of channels for fusing $n$ nontrivial charges to the other vacuum $\ket{2\bar{2}}$.

For general non-abelian models, if both boundaries of a cylinder are of charge type, then the boundary excitation distribution spectrum can always be identified with the bulk entanglement spectrum with a cylinder+disk cut, 
\begin{equation}
\rho_{n,\{n_\alpha,n_\beta\}}=D^{1-L} \rho_{A; n,\{n_\alpha,n_\beta\}}.
\end{equation}
On the left hand side we have the generalized density matrix, while on the right hand side is the reduced density matrix.
The general expression for a cut that splits the $j$-loop and separates a cylindrical subsystem $A$ from a disk $B$ is \footnote{\eqref{eq:CylDiskRho} is the case $L_1=L, L_2=1$ in Ref. \cite{Luo}}, in the language of section \ref{subsec:EntDisk}
\begin{equation}
\label{eq:CylDiskRho}
\rho_A=\sum_{\J} |c_\J|^2 \left\{\oplus_{j\in I} M_{\J j} \frac{d_j}{d_\J}\left[\frac{D}{d_j} P_j (\alpha^{\otimes L})\right]\right\}.
\end{equation}
It again splits into blocks of $\rho_A=\oplus_{n,\{n_\alpha,n_\beta\}}\rho_{A; n,\{n_\alpha,n_\beta\}}.$

\section{Discussion}\label{sec:discussion}
We examine the relaxation of assumptions and the limitation of our methods in this part. Comments on future directions will also follow.

\subsection{General boundary conditions}\label{subsec:future}
The charge and flux boundary cases have been discussed in the above sections, however, for the most general boundary conditions, there are certain subtleties. While the charge boundary always exists, flux boundary may not. This happens in many non-abelian models, for example, the doubled Ising case. So letting the entanglement cut split the $i$-loop does not always give rise to an entanglement spectrum that corresponds to a meaningful boundary excitation distribution spectrum. Additionally, one can have more complicated Frobenius algebras or boundary conditions in addition to these two types. For example, it was shown in Ref. \cite{Juven} that for the $\Z_4$ model, there are three different types of boundary conditions, which will lead to six different combinations and thus energy spectra for the two boundaries of a cylinder. It is unclear how one can realize all the corresponding entanglement spectra from the bulk.
	
The above defects are expected and inevitable because the boundary-bulk correspondence is many-to-one, which already happens in the chiral cases \cite{KitaevE8, LuVishwanath, Chetandim16}. The same bulk theory can share many different boundary theories, even gapless ones \cite{KongGapless}. Consequently, one cannot extract boundary data purely from bulk information. If one hopes to realize all types of boundary conditions from the entanglement point of view, he/she is forced to add boundaries to the whole system when doing the entanglement calculations. For example, one can start from an cylinder with certain boundary conditions, make a specific bipartition so that the boundaries of both subsystems partially coincide with the boundaries of the open surface (e.g., subsystem $A$ is a vertical slit of the cylinder that touches both boundaries of the cylinder, and $B$ is the rest). It was shown in the quantum double and the continuous cases recently \cite{Juven1801, Janet1804,Janet1901}, that the corresponding entanglement entropy is explicitly dependent on boundary conditions of the cylinder. Within this setup, we expect to be able to distinguish all different boundary conditions from the entanglement spectrum. In a similar spirit but from the information-theoretical perspective, Ref. \cite{Kato} argues that if the subsystem $A$ is chosen as the region around the physical boundary, then the entanglement spectrum is equivalent to the spectrum of an edge state living on the boundary. Ref. \cite{Bowen} also makes the subsystem touch the gapped boundaries and is able to obtain boundary-dependent entanglement properties, using the concept of information convex. These strategies, however, unavoidably involves extra input than purely bulk data, and is not chosen in this paper.
	
One future direction would be to look into more exotic boundary conditions, such as those given by fermion condensations \cite{FermionicWan, FermionicJuven}. Fermion condensation is expected to be described by twisted Frobenius algebras \cite{FermionicWan} and may have similar results for the correspondence. In addition, global symmetry will lead to more possibilities for boundary conditions such as those constructed from symmetry extension \cite{WWW, Juven1801}. It would be interesting to extend our discussion to symmetry enriched topological phases and study the bulk-boundary correspondence there.

\subsection{The small constant $\epsilon$}
	We have focused on the case where $\epsilon$ is a small positive constant in the Hamiltonian \eqref{eq:FullHam}.  If one breaks the topological invariance of the boundary theory by varying $\epsilon$ with positions or by adding a generic perturbation, the above correspondence will, in general, be lost. (The case of $\mathbb{Z}_2$ was discussed in Ref. \cite{Vidal}.) But we argue that the correspondence still holds if (i) $\epsilon$ stays as a constant but takes a larger value, or (ii) $\epsilon$ remains small but becomes position-dependent.
	
(i) Previously $\epsilon$ was taken to be small because we would like to focus on the boundary excitations only, and require the bulk to be always in its ground state. Starting from the ground state on a sphere $S^2$, one can go beyond the bulk ground state subspace by creating a pair of quasiparticles $\J$, $\bar{\J}$ and move them to the two disk-shaped subsystems $A$, $\bar{A}$, respectively. The diagonalized reduced density matrix of subsystem $A$ is then $\rho_A=\oplus_j M_{\J j} \frac{D}{d_j} P_j (\alpha^{\otimes L})$. All entries are products of quantum dimensions $d_{i_1}, d_{i_2}, \cdots, d_{i_L}$ (up to some power of $D$), which can again be organized according to the number of nontrivial $i\neq 0$'s as in (3.6), $\rho_A=\oplus_{n,\{n_\alpha\}} \rho_{A; {n,\{n_\alpha\}}}$.
	
Above $A$ is an abstract disk generated from the bipartition. Alternatively, we can look at a physical disk $D^2$ with a charge boundary for convenience. Initially, both the bulk and boundary of the disk are in their ground state. Then one creates a pair of excitations $\J'$, $\bar{\J'}$ in the bulk and move one of them $\J'$ to the boundary, so that the bulk has only one excitation $\bar{\J'}$ left. On the boundary, $\J'$ decomposes according to $\J'\rightarrow \oplus_j M_{\J' j} j$. The degeneracy for the distribution $\{n_\alpha\}$ of boundary excitations is given by a generalization of \eqref{eq:deg-n},
\begin{equation}
\text{deg}_{n,\{n_\alpha\}} (D^2) =\sum_j  M_{\J' j} g_{j; n, \{n_\alpha\}}\frac{L!}{n_0!n_\alpha!\cdots n_\zeta!},	\end{equation}
where $g_{j; n, \{n_\alpha\}}$ is the number of fusion channels for $n_\alpha$ charges $\alpha$, $n_\beta$ charges $\beta$ etc. to fuse to charge $j$. This deg$_{n,\{n_\alpha\}}(D^2)$ is matched with the dimension of the above $\rho_{A;{n,\{n_\alpha\}}}$ when $\J'=\J$. Namely, there is still a correspondence between the energy spectrum of a disk when there exists bulk excitations and the entanglement spectrum of a sphere in its excited state, but this correspondence requires the set of quasiparticles in these two cases to be the same. 
	
(ii) If $\epsilon=\epsilon(n)>0$ is position-dependent, then the boundary energy spectrum becomes disorganized {at finite temperatures}. However, the boundary excitation distribution spectrum is the same as before: $n$ no longer labels the $n$-th excited state, but still labels the excited state with $n$ boundary quasiparticles. {Furthermore, as we are interested in the high temperature limit of the boundary spectrum, the changes to the energy levels are irrelevant.}

One can similarly vary the terms in the bulk Hamiltonian along the entanglement cut. As long as $\epsilon(n)$ is small and positive, the ground state remains the same, and the corresponding entanglement spectrum is the same. Hence the correspondence between bulk entanglement spectrum and the boundary excitation distribution spectrum is robust against such variations.
	
\subsection{The infinite-temperature \& fixed-fugacity limit}
{In the infinite temperature limit, all excitations occur with the same possibility. The only difference comes from the chemical potential, or the fugacity. The importance of fugacity is special for string-net models. For finite group cases, were we in the group element basis as in Kitaev quantum double models, the fugacities should all be trivial. For example, in Ref. \cite{Yuting1901}, the authors observe that the entanglement entropy in Kitaev quantum double models is equal to the thermal entropy of a 1d system at infinite temperature. In the string-net casess, we are using the representation basis instead. This introduces internal degrees of freedom in the open links intersectng the entanglement cut. These degrees of freedom are related to their quantum dimensions: the number of channels for $n_\alpha$ excitations of type $\alpha$ to fuse to vacuum grows asymptotically like the $n_\alpha$-th power of $d_\alpha$ \cite{KPTEE}. Consequently, the quantum dimension $d_\alpha$ can be viewed the asymptotic dimension of the Hilbert space $\mathcal{H}_\alpha$ containing one particle of type $\alpha$ . (This is only ``asymptotic'' because the dimension of Hilbert space is always an integer, while quantum dimensions can be non-integral.)  Then $\prod_{\alpha} d_\alpha^{n_\alpha}$ in \eqref{eq:DM} counts the asymptotic dimension of a tensor product of Hilbert spaces $\otimes_\alpha \mathcal{H}_\alpha^{\otimes n_\alpha}.$}

{The identification of quantum dimension as fugacity gives a hint to the statistical mechanical theory of anyonic systems. We leave the full understanding to future work.}

\begin{acknowledgments}
Zhu-Xi thanks Ling-Yan Hung and Bowen Shi for helpful discussions.
\end{acknowledgments}
	
\flushbottom

	
\end{document}